\documentclass[%
aip,
jcp,
]{revtex4-1}
\usepackage{graphicx}
\usepackage{multirow,textcomp}
\usepackage{amsmath,mathrsfs,amsthm,amsfonts,mathtools}
\usepackage{accents}
\usepackage{comment}
\usepackage{color}
\usepackage{xcolor}
\usepackage{longtable}
\usepackage{verbatim}
\usepackage{natbib}
\usepackage{hyperref}
\usepackage{colortbl}
\usepackage{multirow}
\usepackage[latin9]{inputenc}
\begin{document}
\title{A lower scaling four-component relativistic coupled cluster method based on natural spinors}
\author{Somesh Chamoli}
\author{Kshitijkumar Surjuse}
\affiliation{Department of Chemistry, Indian Institute of Technology Bombay, Powai, Mumbai 400076, India}
\author{Malaya K. Nayak}
\email{mknayak@barc.gov.in }
\affiliation{Theoretical Chemistry Section, Bhabha Atomic Research Centre, Trombay, Mumbai 400085, India}
\affiliation{Homi Bhabha National Institute, BARC Training School Complex, Anushakti Nagar, Mumbai 400094, India}
\author{Achintya Kumar Dutta}
\email{achintya@chem.iitb.ac.in}
\affiliation{Department of Chemistry, Indian Institute of Technology Bombay, Powai, Mumbai 400076, India}

\begin{abstract}
    We present the theory, implementation, and benchmark results for a frozen natural spinors-based lower scaling four-component relativistic coupled cluster method. The natural spinors are obtained by diagonalizing the one-body reduced density matrix from a relativistic MP2 calculation based on four-component Dirac-Coulomb Hamiltonian. The correlation energy in the coupled cluster method converges more rapidly with respect to the size of the virtual space in the frozen natural spinor basis than that observed in the standard canonical spinors obtained from the Dirac-Hartree-Fock calculation. The convergence of properties is not smooth in the frozen natural spinor basis. However, the inclusion of the perturbative correction smoothens the convergence of the properties with respect to the size of the virtual space in the frozen natural spinor basis and greatly reduces the truncation errors for both energy and properties calculations. The accuracy of the frozen natural spinor based coupled cluster methods can be controlled by a single threshold and is a black box to use.
\end{abstract}

\maketitle
\section{Introduction}
Accurate simulation of energy and properties of heavy elements requires inclusion of both relativistic effect and electron correlation in the electronic structure calculation. The relativistic coupled cluster method, based on four or two-component Dirac-Coulomb Hamiltonian, has emerged as the method of choice for quantum chemical calculation of heavy elements due to its systematic improvable nature\cite{Kaldor2002RELATIVISTICELEMENTS,Liu2021RelativisticMethods}. Efficient implementation of relativistic coupled cluster methods for energy, properties and excited state calculation is described in the literature\cite{doi:10.1063/1.3518712,doi:10.1063/1.4966643,PhysRevA.97.022512,PhysRevA.91.030503,PhysRevA.90.062501,doi:10.1063/1.5053846}. The extension to multi-reference variants of the coupled cluster method which can effectively treat quasi-degenerate electronic states has also been achieved\cite{doi:10.1063/1.4962911,Eliav2010RelativisticApproach,Fleig2007AHBr,doi:10.1063/1.1323258}.\\
However, one of the major bottlenecks of all these implementations is the inherent high computation cost of the coupled cluster method. The commonly used singles-doubles (CCSD) truncation of the cluster operator has formal computation scaling of $O(N^{6})$ power of the basis set, whereas the inclusion of partial triples correction in the form of CCSD(T) increases the computational scaling to $O(N^{7})$ power of the basis set. The computation becomes more demanding  for the relativistic coupled cluster calculation on heavy elements due to the increase in the number of  electrons, larger dimension of the basis set and and the switch from real to complex algebra.  Attempts have been made to reduce the computational cost of the relativistic calculation by using density fitting approximation for the two electron integrals\cite{HELMICHPARIS201938}, or Laplace transform techniques\cite{doi:10.1063/1.4955106}. However, practical implementations has only been achieved for second order M{\o}ller-Plesset perturbation theory.  Alternatively computational speed up can be achieved by using computational programs which can scale on multiple cores. Lucas Visscher and co-workers have reported a massively parallel implementation of relativistic coupled cluster theory for  GPU-accelerated computing architectures\cite{doi:10.1021/acs.jctc.1c00260}. We want to use a more compact set of basis functions in the form of natural spinors to reduce the computational cost of relativistic coupled cluster calculations. \\
In the non-relativistic quantum chemistry, the natural orbitals are defined as the eigen-functions of a correlated one-body reduced density matrix\cite{PhysRev.97.1474}. Natural orbitals give a more compact description of the orbital space than that obtained from the canonical Hartree-Fock orbitals. One can use natural orbitals generated from the approximate correlation method (generally MP2) to truncate the dimension of the basis set in the coupled cluster calculations. The computational cost of coupled cluster calculations can be significantly reduced using the truncated natural orbital basis \cite{taube2005frozen,doi:10.1063/1.2902285,Neese2009}.  We propose to use a similar approach in the relativistic domain using natural spinors  which are the eigen-functions of the one-body reduced density matrix of a spin-orbit coupled wave-function obtained from a relativistic electron correlation theory. The idea of natural spinors have been explored before also\cite{doi:10.1063/1.3592780,doi:10.1021/ct300205r}, but mostly as an interpretive tool. In this work we explore the suitability of natural spinors for the reduction of computational cost of relativistic coupled cluster calculations.

\section{Theory and computational details}
\subsection{Relativistic Coupled Cluster Method}
The primary approach to solve any many-body problem in non relativistic quantum chemistry is to take each electron under an average field of other electrons. This approach is also known as Hartree-Fock method. One can use the Dirac one-electron Hamiltonian of relativistic theory in Hartree-Fock method, and this method can be referred to as Dirac-Hartree-Fock (DHF) or relativistic self consistent field method\cite{Reiher2015}. With Dirac-Coulomb Hamiltonian, the DHF equation can be written in its matrix form  as  
\begin{equation}
\label{eq:1}
\begin{bmatrix}
\hat{V}+\hat{J}-\hat{K} & \hspace{2cm}c(\sigma\cdot\hat{P})-\hat{K} \\
c(\sigma\cdot\hat{P})-\hat{K} &\hspace{2cm} \hat{V}-2mc^{2}+\hat{J}-\hat{K} 
\end{bmatrix}\begin{bmatrix}
\psi^{L}  \\
\psi^{S}
\end{bmatrix}=E\begin{bmatrix}
\psi^{L}  \\
\psi^{S}
\end{bmatrix}\hspace{2cm}
\end{equation}

where $\psi^{L}$ and $\psi^{S}$ denotes large and small components of 4-spinor $\psi$ with each component itself having a form of 2-spinor.
\begin{equation}
\label{eq:2}
\psi^{L}=\begin{bmatrix}
\phi_{\alpha}^{L}  \\
\phi_{\beta}^{L}
\end{bmatrix}\hspace{1cm}and\hspace{1cm} \psi^{S}=\begin{bmatrix}
\phi_{\alpha}^{S}  \\
\phi_{\beta}^{S}
\end{bmatrix}
\end{equation}
The $\hat{V}$ in eq (1) denotes nuclear-electron interaction, $\hat{P}$ represent momentum operator, $\sigma$ are pauli spin matrices,  m is mass of the electron and c is the speed of light. The direct electron-electron interaction and exchange operators are defined as
\begin{equation}
\label{eq:3}
\hat{J}=\sum_{i}^{N}\int{\psi_{i}^{\dagger}(2)\hat{g}(1,2)\psi_{i}(2)dr_{2}}
\end{equation}
\begin{equation}
\label{eq:4}
\hat{K} \psi_{j}^{X}(1)=\sum_{i}^{N} \sum_{Y}^{L, S} \int \psi_{i}^{\dagger Y}(2) \hat{g}(1,2) \psi_{j}^{X}(2) d r_{2} \psi_{i}^{Y}(1)
\end{equation}

where summation index i is used for occupied spinors and N is the total number of electrons. The instantaneous Coulomb operator $\hat{g}(1,2)$ which give rise to both direct and exchange terms is defined as:
\begin{equation}
\label{eq:5}
\hat{g}(1,2)=\frac{e^{2}}{4\pi\epsilon_{0}}\frac{1}{r_{ij}}
\end{equation}
In the above operator only electrostatic interactions between electrons are considered but one can add higher order relativistic corrections to the Hamiltonian. The two-electron operator with Breit correction is written as:
\begin{equation}
\label{eq:6}
\hat{g}_{B}(1,2)=-\frac{e^{2}}{4\pi\epsilon_{0}r_{ij}}{(\alpha_{i}\cdot\alpha_{j})}+\frac{1}{2}\left[{\frac{e^{2}}{4\pi\epsilon_{0}r_{ij}}{(\alpha_{i}\cdot\alpha_{j})}}-\frac{e^{2}}{4\pi\epsilon_{0}}\frac{{(r_{ij}\cdot\alpha_{i})(r_{ij}\cdot\alpha_{j})}}{{r_{ij}^{3}}}\right]
\end{equation}
with $\alpha$ matrices for an individual electron given by
\begin{equation}
\label{eq:7}
\alpha=\begin{bmatrix}
\textbf{0} & \sigma_{J}  \\
\sigma_{J} & \textbf{0}
\end{bmatrix}\hspace{1.5cm}[J=x,y,z]
\end{equation}
However, in the present study we have restricted the two-electron operator only to the electrostatic interactions. The straight forward use of the relativistic Dirac-Coulomb Hamiltonian is not possible due to the unbounded nature of the Hamiltonian which results in negative energy solutions, in addition  with the positive energy solutions.
Since we are only interested in the positive spectrum of the energy, one can define a Hamiltonian that restricts the electron to stay only in the positive region of the spectrum. Such Hamiltonians are known as no-pair Hamiltonian as they hinder the pair creation processes\cite{doi:10.1063/1.4959452}. No-pair Hamiltonian can be constructed using the projection operators.
\begin{equation}
\label{eq:8}
\hat{H}^{+}=\sum_{i}^{N}\Lambda^{+}\hat{h}^{D}(i)\Lambda^{+}+\sum_{i<j}^{N}\Lambda^{+}\hat{g}(i,j)\Lambda^{+}
\end{equation}
$\Lambda^{+}$ in equation (8) stands for projection operator. $\hat{H}^{+}$ denotes no-pair Hamiltonian and $\hat{h}^{D}(i)$ is one electron Dirac Hamiltonian. 

We have an advantage of using spin-symmetry in non-relativistic theory, due to which the spatial and spin coordinates can be treated separately, resulting in the reduction of computational cost. The spin symmetry is not present in relativistic cases as it gets lost due to the spin-orbit coupling. However, reduction in the computational cost can be achieved with the use of the time reversal symmetry\cite{landau2013course}. Under the time reversal symmetry, the fermionic states are doubly degenerate in the absence of external vector potential and the combination of such states is referred to as Kramers pair. The two degenerate states $\psi$ and $\Bar{\psi}$  can be related to each other by Kramers time-reversal operator $\hat{\kappa}$.
\begin{equation}
\label{eq:9}
\hat{\kappa}\psi=\Bar{\psi}
\end{equation}
where the expression for $\hat{\kappa}$ is
\begin{equation}
\label{eq:10}
\hat{\kappa}=-i\begin{bmatrix}
\sigma_{y} & 0  \\
0 & \sigma_{y} 
\end{bmatrix}K_{0}
\end{equation}
with $K_{0}$ as complex conjugation operator.

In the coupled cluster approach\cite{shavitt_bartlett_2009} the wave function has an `exponential ansatz' given by
\begin{equation}
\label{eq:11}
\Psi_{cc}=e^{\hat{T}}\Phi_{0}
\end{equation}
where $\Phi_{0}$ is a reference determinant and $\hat{T}$ stands for cluster operator, which produces different excited determinants form the reference determinant ($\Phi_{0}$).
\begin{equation}
\label{eq:12}
\hat{T}=\hat{T_{1}}+\hat{T_{2}}+....+\hat{T_{n}}
\end{equation}
with 
\begin{equation}
\label{eq:13}
\hat{T_{1}}=\sum_{ia}{t_{i}^{a}}a_{a}^{\dagger}a_{i}
\end{equation}
\begin{equation}
\label{eq:14}
\hat{T_{2}}=\frac{1}{4}\sum_{ijab}{t_{ij}^{ab}}a_{a}^{\dagger}a_{b}^{\dagger}a_{j}a_{i}
\end{equation}
\begin{equation}
\label{eq:15}
\hat{T_{n}}=\left(\frac{1}{n!}\right)^{2}\sum_{ij...ab...}^{n}{t_{ij...}^{ab...}}a_{a}^{\dagger}a_{b}^{\dagger}...\hspace{0.2cm}a_{j}a_{i}...
\end{equation}
as one-spinor, two-spinor .... n-spinor cluster operators. Here $\left(t_{i}^{a}\hspace{0.1cm}t_{ij}^{ab}\hspace{0.1cm}....\hspace{0.1cm}t_{ij...}^{ab...}\right)$ are cluster amplitudes, indices $\left(i\hspace{0.1cm}j\hspace{0.1cm}k\hspace{0.1cm}...\right)$ denotes occupied spinors and $\left(a\hspace{0.1cm}b\hspace{0.1cm}c\hspace{0.1cm}...\right)$ denotes virtual spinors of a reference determinant. Truncation of cluster operator only up to 1-body and 2-body excitations leads to the commonly used coupled cluster singles-doubles approximation (CCSD).  The most expensive step in the CCSD calculation has a formal scaling of $O (N_{O}^{2}N_{V}^{4})$ where $N_{O}$ denotes the number of occupied spinors and $N_{V}$ denotes the number of virtual spinors.

One can derive the CC equations for the energy and amplitudes using projection technique
\begin{equation}
\label{eq:16}
\langle \Phi_{0}|\Bar{H}|\Phi_{0}\rangle=E\
\end{equation}
\begin{equation}
\label{eq:17}
\langle \Phi_{ij...}^{ab...}|\Bar{H}|\Phi_{0}\rangle=0
\end{equation}
where $\Bar{H}=e^{-\hat{T}}\hat{H}e^{\hat{T}}$ is the coupled cluster similarity transformed Hamiltonian. The advantage due to the Campbell-Baker-Hausdroff expansion in similarity transformed Hamiltonian is that now the CC equations truncates naturally beyond the fourth power of $\hat{T}$.
\begin{equation}
\label{eq:18}
\bar{H}=\hat{H}+[\hat{H}\hspace{0.1cm}\hat{T}]+\frac{1}{2!}\left[\left[\hat{H}\hspace{0.15cm}\hat{T}\right]\hspace{0.15cm}\hat{T}\right]+\frac{1}{3!}\left[\left[\left[\hat{H}\hspace{0.15cm}\hat{T}\right]\hspace{0.15cm}\hat{T}\right]\hspace{0.15cm}\hat{T}\right]+....
\end{equation}
The expression for $\bar{H}$ can be further simplified using normal ordered Hamiltonian.
\begin{equation}
\label{eq:19}
\bar{H}=\left(\hat{H_{N}}+\hat{H_{N}}\hspace{0.1cm}\hat{T_{1}}+\hat{H_{N}}\hspace{0.1cm}\hat{T_{2}}+\frac{1}{2}\hat{H_{N}}\hspace{0.1cm}\hat{T_{1}^{2}}+\hat{H_{N}}\hspace{0.1cm}\hat{T_{1}}\hspace{0.1cm}\hat{T_{2}}+....\right)_{c}
\end{equation}
The subscript c in equation (19) denotes that only the connected terms in $\bar{H}$ i.e. only those terms in which $\hat{T}$ are explicitly connected to the normal ordered Hamiltonian, survives in the final expression. 

The CCSD method often fails to provide quantitative accuracy and the consideration of higher excitation (for e.g. triple or higher) becomes crucial. Unfortunately the full inclusion of $\hat{T_{3}}$ cluster operator 
 \begin{equation}
\label{eq:20}
\hat{T}=\hat{T_{1}}+\hat{T_{2}}+\hat{T_{3}}
\end{equation}
(i.e. the full CCSDT method\cite{doi:10.1063/1.452353,SCUSERIA1988382}) leads to a very high computational cost, which scales as O($N_{O}^{3}N_{V}^{5}$) power of the basis set. However, the inclusion of a part of the three-body excitation operators (using perturbational\cite{doi:10.1063/1.449067} or active space argument\cite{doi:10.1080/00268976.2010.522608}) is often sufficient to get desired accuracy. Among the various perturbative triples correction schemes available, the  CCSD(T) method \cite{RAGHAVACHARI1989479} gives the best compromise between computational cost and accuracy and is considered as the `gold standard' method in quantum chemistry.  In the  CCSD(T) method, the triples energy correction is calculated non-iteratively using the converged CCSD amplitudes and is defined as follows
\begin{equation}
\label{eq:21}
E_{(T)}=\frac{1}{36}\sum_{abc}\sum_{ijk}t_{ijk}^{abc}(c)\hspace{0.1cm}D_{ijk}^{abc}\left[t_{ijk}^{abc}(c)+t_{ijk}^{abc}(d)\right]
\end{equation}
where 

\begin{equation}
\label{eq:22}
D_{ijk}^{abc}=f_{ii}+f_{jj}+f_{kk}-f_{aa}-f_{bb}-f_{cc}
\end{equation}
is expressed in terms of diagonal Fock matrix elements. The disconnected and connected triple excitation amplitudes can be defined through the relations
\begin{equation}
\label{eq:23}
D_{ijk}^{abc}\hspace{0.1cm}t_{ijk}^{abc}(d)=P(i/jk)\hspace{0.1cm}P(a/bc)\hspace{0.1cm}t_{i}^{a}\hspace{0.1cm}{\langle jk||bc \rangle}
\end{equation}
\begin{equation}
\label{eq:24}
D_{ijk}^{abc}\hspace{0.1cm}t_{ijk}^{abc}(c)=P(i/jk)\hspace{0.1cm}P(a/bc)\hspace{0.1cm}\left[\sum_{e}t_{jk}^{ae}\hspace{0.1cm}{\langle ei||bc\rangle}-\sum_{m}t_{im}^{bc}\hspace{0.1cm}{\langle ma||jk\rangle}\right]
\end{equation}
with $P(i/jk)$ and $P(a/bc)$ as 3-index permutation operator whose operation on any arbitrary function $X$ can be defined as 
\begin{equation}
\label{eq:25}
P(i/jk)\hspace{0.1cm}X(ijk)=X(ijk)-X(jik)-X(kji)
\end{equation}
The amplitudes $t_{i}^{a}$ and $t_{ij}^{ab}$ are converged CCSD amplitudes and ${\langle ab||ij\rangle}$ are antisymmetrized two electron integrals. The total CCSD(T) energy is given by
\begin{equation}
\label{eq:26}
E=E_{CCSD}+E_{(T)}
\end{equation}
with $E_{CCSD}$ as solution of equation (16). The implementation of the CCSD(T) method based on a four-component Dirac-Coulomb Hamiltonian has already been reported in the literature\cite{doi:10.1063/1.472655,doi:10.1021/acs.jctc.1c00260}.
The most time consuming step in the CCSD(T) scales as O($N_{O}^{3}N_{V}^{4}$) power of the basis set.

\subsection{Natural Spinors}
Following L\"{o}wdin\cite{PhysRev.97.1474}, the natural orbitals are defined as an eigen states of one-body reduced density matrix and can be generated by diagonalizing the one-body reduced density matrix obtained from a correlation calculation. MBPT(2) is generally used for the generation of natural orbitals due to its favourable scaling. Various non-relativistic post HF methods can be cast into a relativistic post DHF methods, if we stick to the no-pair approximation. Therefore, natural spinors can be determined in the same analogy as natural orbitals in the non relativistic regime:
First, a DHF calculation with no pair approximation is performed which yields positive occupied and positive  virtual molecular spinors only. Now the virtual-virtual block of the one-body reduced density matrix can be constructed using MBPT(2) method as: 
\begin{equation}
\label{eq:27}
 D_{ab}=\sum_{cij}^{} {\frac{\langle ac||ij \rangle \hspace{0.1cm}\langle ij||bc \rangle}{\varepsilon_{ij}^{ac} \hspace{0.2cm}\varepsilon_{ij}^{bc}}}
\end{equation}
where 
\begin{equation}
\label{eq:28}
\varepsilon_{ij}^{ac}=\varepsilon_{i}+\varepsilon_{j}-\varepsilon_{a}-\varepsilon_{c}
\end{equation}
\begin{equation}
\label{eq:29}
\varepsilon_{ij}^{bc}=\varepsilon_{i}+\varepsilon_{j}-\varepsilon_{b}-\varepsilon_{c}
\end{equation}
with $\varepsilon_{i}$, $\varepsilon_{j}$, $\varepsilon_{a}$, $\varepsilon_{b}$, and $\varepsilon_{c}$ as DHF orbital energies and $\langle ac||ij \rangle$ and $\langle ij||bc \rangle$ are antisymmetrized two electron integrals. This virtual-virtual block of the one body reduced density matrix is then diagonalized.
\begin{equation}
\label{eq:30}
DV=Vn
\end{equation}
Here V are virtual natural spinors and the eigenvalues n obtained are occupation numbers corresponding to virtual natural spinors. We can sort these virtual spinors on the basis of their importance towards total correlation energy using the occupation info. As a result, the virtual spinor space can be truncated by preserving only the virtual natural spinors whose occupation number are above a predetermined threshold or cutoff. The Fock matrix's virtual-virtual block are then transformed into the truncated natural spinor basis. 
\begin{equation}
\label{eq:31}
F_{vv}^{NS}=\tilde{V}^{\dagger}F_{vv}\tilde{V}
\end{equation}
where $\tilde{V}$ are truncated natural spinors. $F_{vv}$ is the virtual-virtual block of initial Fock matrix and $F_{vv}^{NS}$ indicates virtual-virtual block of the Fock matrix in truncated virtual natural spinor basis. The canonical virtual natural spinors and their accompanying orbital energies are obtained by diagonalizing this fock matrix in the new natural spinor basis. 
\begin{equation}
\label{eq:32}
F_{vv}^{NS}\tilde{Z}=\tilde{Z}\epsilon
\end{equation}
In equation (32), $\tilde{Z}$ represents canonical virtual natural spinors and $\epsilon$ denotes their orbital energies. 
The matrix
\begin{equation}  
U=\tilde{Z}\tilde{V}
\protect\label{eq:EOM4}
\end{equation}
denotes the transformation from the canonical DHF virtual space to canonical natural virtual spinor space. As a result, our basis set is composed of canonical DHF occupied spinors and canonical natural virtual spinors. Since our occupied spinors are frozen in their original DHF spinor form, so this approach is referred to as the frozen natural spinor (FNS) method. Higher order correlation calculations can be performed in the FNS basis at a lower computational cost and hopefully without significant errors. The FNS-CCSD method scales as O($N_{O}^{2}(N_{V}-N_{FV})^{4}$), where $N_{FV}$ is the number of discarded virtual spinors. The FNS-CCSD(T) method  scales as O($N_{O}^{3}(N_{V}-N_{FV})^{4}$) power of the basis set.
\\One can perform a perturbative correction for the truncated virtual space by approximating the $\Delta E_{CCSD/CCSD(T)}$ as $\Delta E_{MP2}$. Where 
\begin{equation} 
\Delta E_{CCSD/CCSD(T)}=E^{FNS}_{CCSD/CCSD(T)}-E^{Canonical}_{CCSD/CCSD(T)}
\end{equation}
and
\begin{equation} 
\Delta E_{MP2}=E^{FNS}_{MP2}-E^{Canonical}_{MP2}
\end{equation}
The perturbative correction has been shown to greatly enhance the accuracy of natural orbital based non-relativistic coupled cluster methods\cite{Taube2005FrozenTheory,Neese2009}. 

The FNS-CCSD and FNS-CCSD(T) method as described above is implemented in our in-house quantum chemistry software package BAGH\cite{BAGH}. The BAGH is written primarily in python, with computational bottleneck parts in Cython and Fortran. BAGH can perform both relativistic and non-relativistic coupled cluster and many-body perturbation theory calculations for ground and excited states. BAGH relies on other quantum chemistry software packages to generate the one and two-electron integrals and the Hartree-Fock/Dirac-Hartree-Fock orbitals. The BAGH is currently interfaced with three different software packages: PySCF\cite{https://doi.org/10.1002/wcms.1340}, GAMESS\cite{GAMESS} and DIRAC\cite{DIRAC21}. The four-component FNS-CCSD and FNS-CCSD(T) calculations are performed using DIRAC19 interface of BAGH. The Fock matrix and integrals are transformed to the FNS basis using the transformation matrix $U$ (equation 33). The accuracy of our FNS-CCSD and FNS-CCSD(T) implementation has been checked by comparing the results at very tight truncation thresholds with the canonical CCSD and CCSD(T) implementation of DIRAC19. The dipole moment, bond length and frequencies are calculated using numerical derivative technique.
\section{Result and discussion}
To check the performance of the FNS-CCSD and FNS-CCSD(T) method, we have chosen the example of hydrogen halide series (HX, X = F, Cl, Br, I) and coinage metal hydrides (YH, Y = Cu, Ag and Au). The aug-cc-pVTZ basis set were used for H, F and Cl. The dyall.acv3z basis set is used for Br, I, and dyall.ae3z basis set were used for Cu, Ag and Au. The uncontracted version of the basis set and frozen core approximation is used for the calculations (except for HF and HCl). The number of frozen occupied orbital for each molecule is denoted in the Table S1.
\subsection{Correlation energy}
\begin{figure}[h!]
    \centering
    \includegraphics[scale=0.5]{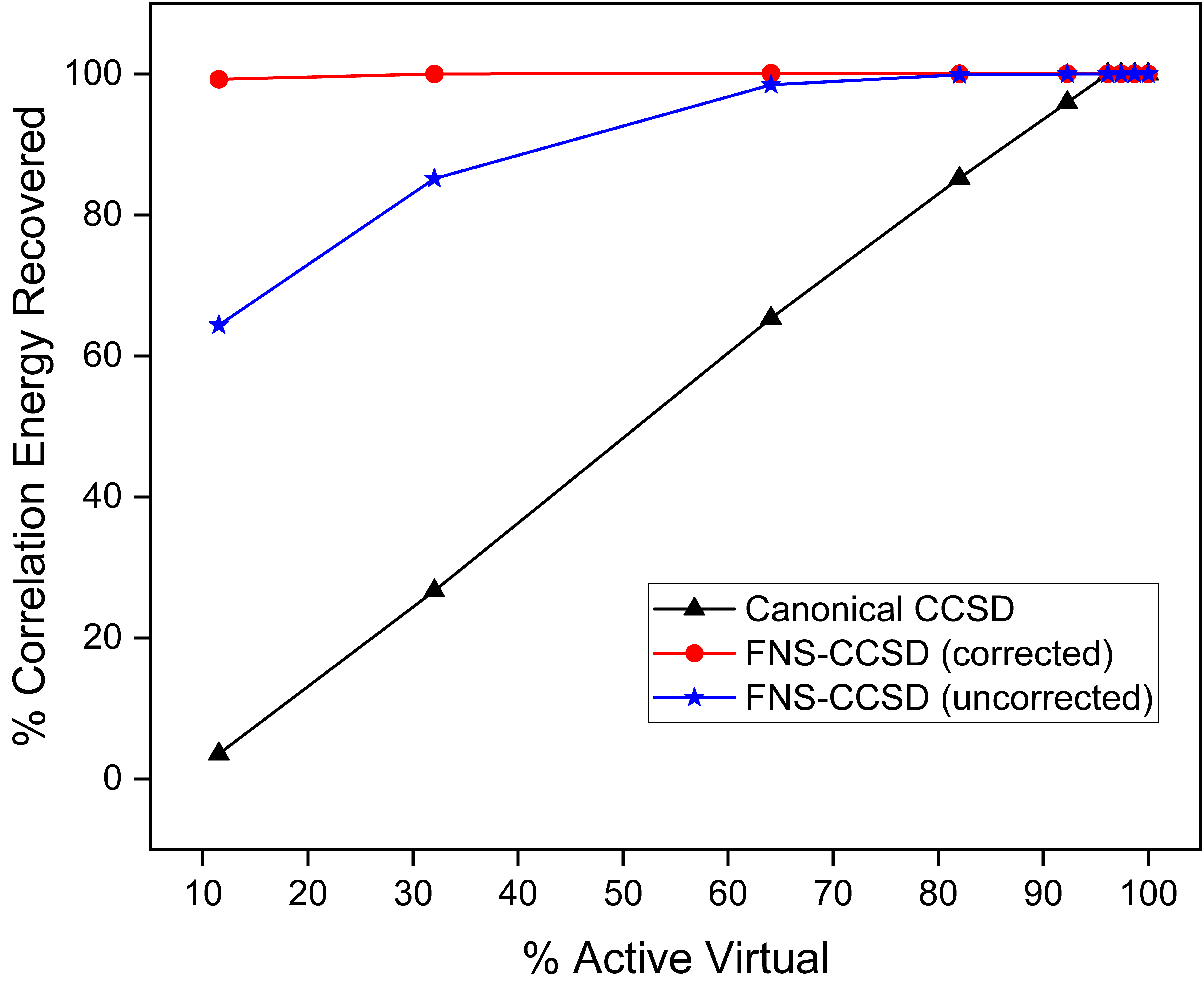}\\
 \caption{ The convergence of the correlation energy in four-component relativistic CCSD method with respect to the size of the virtual space in canonical and FNS basis for HF.}
    \label{fig:my_label1}
\end{figure}

The target accuracy of the natural orbital in non-relativistic coupled cluster method\cite{Neesecepa2009} is 
to recover 99.9\% of the correlation energy. Within the framework of the four-component relativistic CCSD and CCSD(T), one should aim for similar level of accuracy. Figure \ref{fig:my_label1} presents the convergence of CCSD correlation energy with respect to the truncated virtual space in the canonical and FNS basis for HF. One can see that the correlation energy converges more rapidly with respect to the size of the virtual space in the FNS basis than that in canonical virtual basis. The inclusion of the perturbative correction further accelerates the convergence. Similar trend is observed for the FNS-CCSD(T) method (see Figure S1).

\begin{figure}[h!]
    \centering
    \includegraphics[scale=0.5]{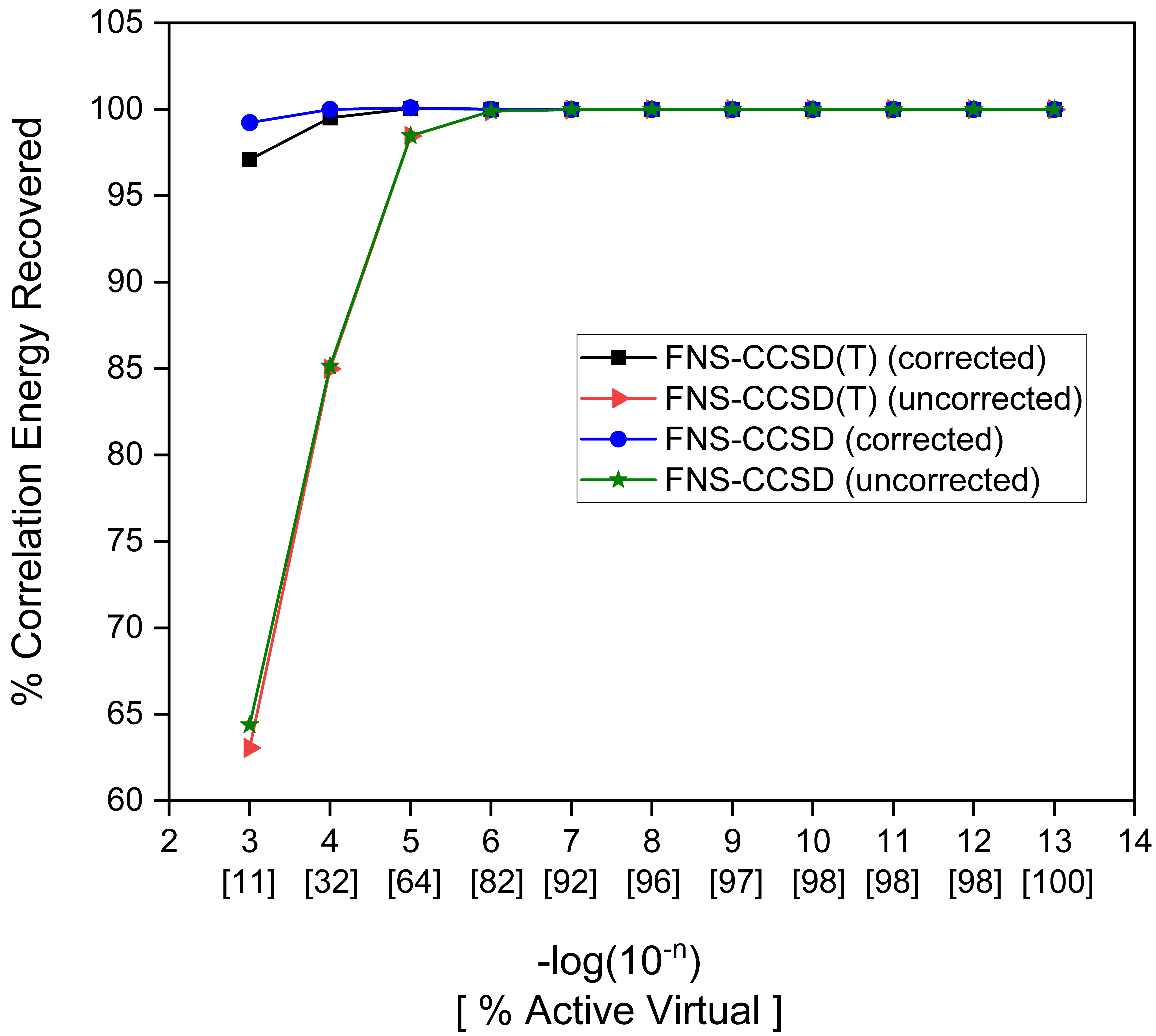}\\
 \caption{ The convergence of FNS-CCSD and FNS-CCSD(T) correlation energies with respect to the FNS truncation threshold ($10^{-n}$) for HF. The percentage of the kept virtual space is provided in the brackets. }
    \label{fig:my_label2}
\end{figure}

It is important to decide a suitable criteria for truncating the natural spinor virtual space. The criteria should be consistent across the molecules, black box to use and equally applicable for energies and various properties. In the original implementation paper of frozen natural orbital based non-relativistic  CCSD by Bartlett and co-workers\cite{Taube2005FrozenTheory}, the percentage of the virtual orbitals were used  as the truncation criteria.  However, the size of the virtual space required to recover a predetermined percentage of the correlation energy in natural orbital/spinor basis is not constant and the occupation of the natural orbital is a more justified criteria to decide the subset of the natural orbitals which can be safely dropped without any significant error in the correlation energy\cite{landau2010frozen,Neese2009}. Therefore, we have used the occupation number of the natural spinor as the criteria for the truncation of the virtual space in the FNS based coupled cluster method. Figure \ref{fig:my_label2} presents the convergence of FNS-CCSD and FNS-CCSD(T) correlation energies with respect to the occupation number for HF. It can be seen that the correlation energies in the FNS-CCSD and FNS-CCSD(T) methods show very similar convergence behaviour with respect to the natural spinor occupation threshold. One need to keep virtual natural spinors up to $10^{-6}$ occupation number to recover more than 99.9 \% of the correlation energy in the uncorrected FNS-CCSD and FNS-CCSD(T) method. One order of magnitude smaller threshold can be used when MP2 perturbative correction is added. In order to make the FNS-CCSD and FNS-CCSD(T) methods a black box one, the same threshold has to work across the molecules. Table \ref{table:1} presents the correlation energies for the hydrogen halides and coinage metal hydrides in FNS-CCSD and FNS-CCSD(T) methods, with an occupation threshold of $10^{-5}$. It can be seen that the occupation threshold of $10^{-5}$ constantly recovers more than 99.9 \% of the correlation energy across the molecules for FNS-CCSD and FNS-CCSD(T) method augmented with MP2 correction. The magnitude of the MP2 correction varies widely from molecule to molecule and shows a range of  0.04 \% to  4.78\%. The uncorrected FNS-CCSD and FNS-CCSD(T) correlation energies are systematically underestimated as compared to the canonical CCSD and CCSD(T) correlation energies. The MP2 corrected  FNS-CCSD and FNS-CCSD(T) correlation energies can be both underestimated and overestimated and the trend varies from molecule to molecule, the consequence of which on property calculation needs to be benchmarked.
The percentage of the kept virtual spinors in the FNS basis also changes from molecule to molecule and does not follow any particular trend. The number of kept virtual spinors in HBr is 178 which is slightly higher than that in HI, although the total dimension of the virtual space is larger in the latter.  The smaller number of kept virtual natural spinors actually recovers a higher percentage (99.32/99.3) of the CCSD/CCSD(T) correlation energy in HI than that observed in HBr (99.30/99.02), when MP2 correction is not considered.
\subsection{Dipole Moment}
\begin{figure}[h!]
    \centering
    \includegraphics[scale=0.5]{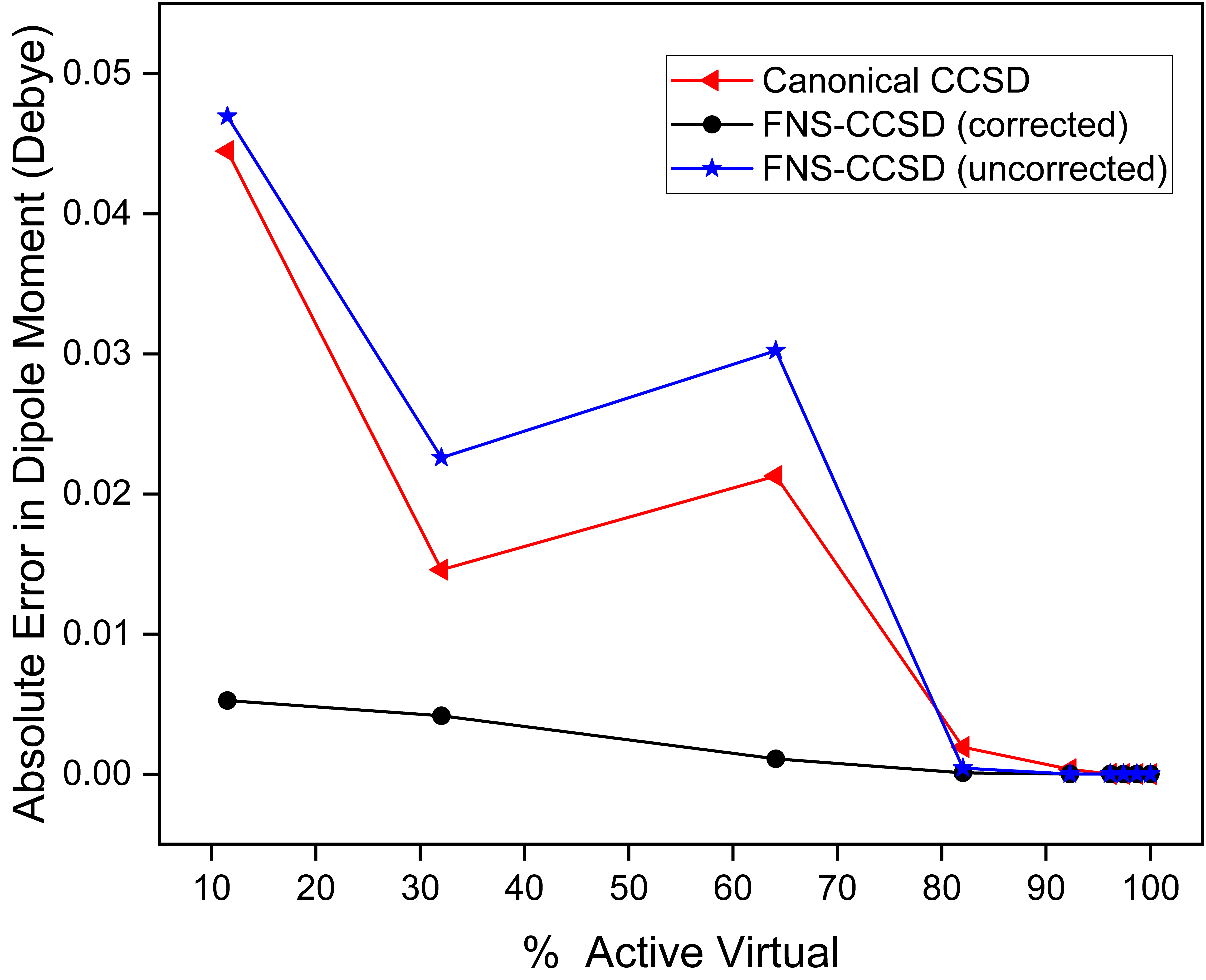}\\
 \caption{The convergence of the four-component relativistic CCSD dipole moment with respect to the size of the virtual space in canonical and FNS basis for HF.}
    \label{fig:my_label3}
\end{figure}

The dipole moment is one of the important one-electron properties which can be easily calculated using the finite-field approximation. The analytic first derivative for CCSD and CCSD(T) methods are available in BAGH\cite{BAGH}. However one need to make the coupled cluster energy functional stationary with respect to the truncated virtual spinors, in addition to making them stationary with respect to the external perturbation\cite{Taube2008FrozenNitroethane}, for analytic calculation of one-electron properties. The dipole moment values presented in this manuscript are calculated using the finite-field approach by adding a positive and negative electric field of 0.0005 a.u. along the molecular axis in the DHF calculation.
\\Figure \ref{fig:my_label3} presents the convergence of CCSD dipole moment  with respect to the size of the virtual space in canonical and FNS basis for HF. One can see that the convergence of the dipole moment with respect to the size of the virtual space is not monotonous in both canonical and FNS basis. The convergence of dipole moment is actually slower in the FNS basis than that in the canonical basis, which is just opposite to the trend observed for energy in Figure \ref{fig:my_label1}. Similar behaviour is observed for FNS-CCSD(T) dipole moment (see Figure S2).

\begin{figure}[h!]
    \centering
    \includegraphics[scale=0.5]{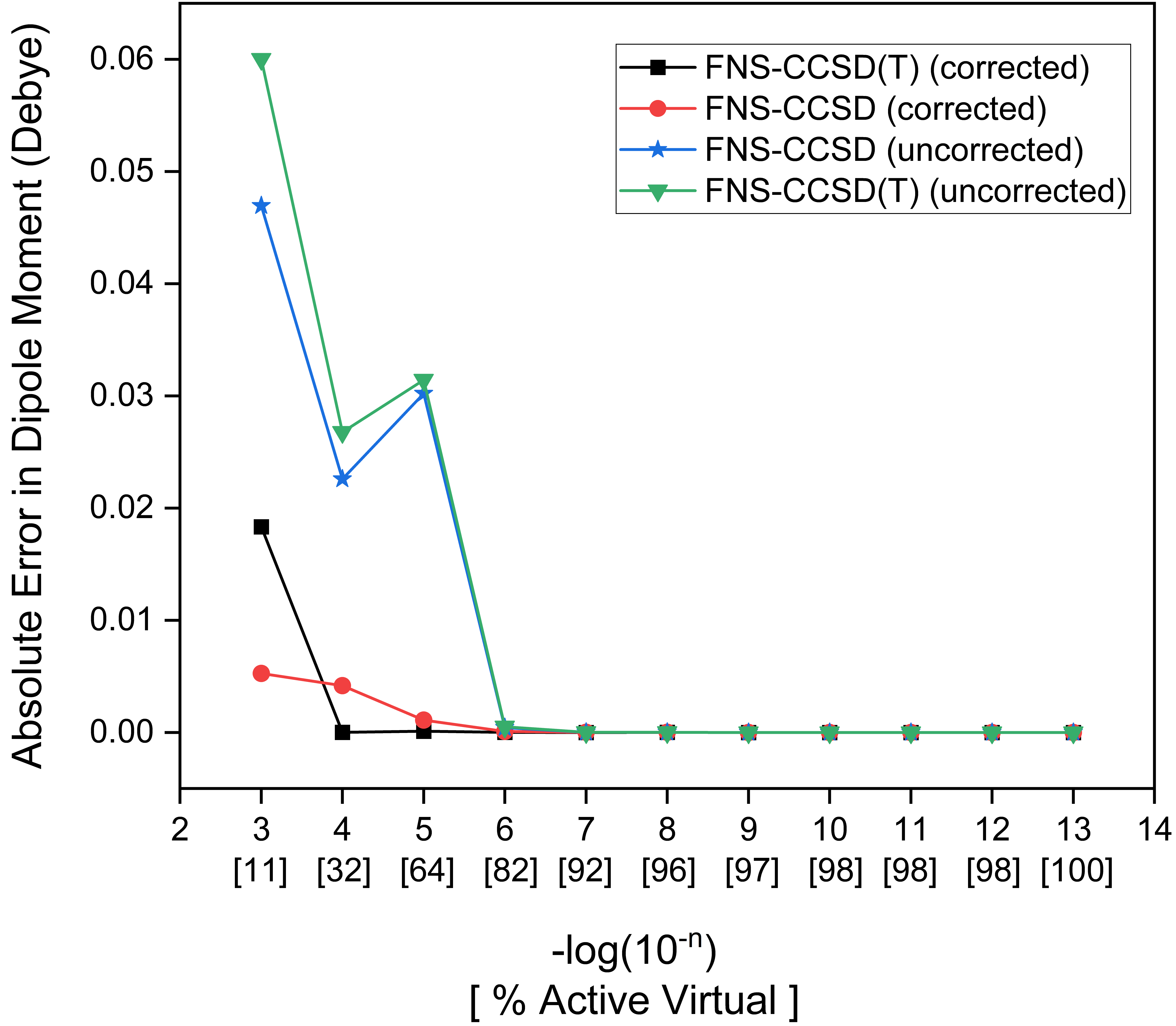}\\
 \caption{  The convergence of FNS-CCSD and FNS-CCSD(T) dipole moment with respect to the FNS truncation threshold ($10^{-n}$) for HF. The percentage of the kept virtual space is provided in the brackets.}
    \label{fig:my_label4}
\end{figure}

However, the trend is consistent with that observed in non-relativistic CCSD based on the frozen natural orbitals for response properties\cite{Kumar2017FrozenTheory}. The source of the error in the response properties for the FNS based coupled cluster method is the ``fundamental dichotomy" between the nature of electron correlation and field-response properties\cite{Crawford2019Reduced-scalingOpportunities}. 
The accuracy of field-response properties strongly depend upon the number of diffuse functions\cite{Woon1998GaussianProperties} in the basis set. The correlation energy, on the other hand, depends primarily on the higher angular momentum functions in the basis set\cite{Dunning1998GaussianHydrogen}. Following the argument as in the case of their non-relativistic counterparts\cite{Crawford2019Reduced-scalingOpportunities} for field-response properties, one can argue that the virtual natural spinors with the highest diffuse character will have the lowest occupation number and will be removed first during the truncation of the virtual space in the FNS procedure. On the other hand, the canonical virtual spinors with the highest diffuse character will correspond to the lowest orbital energies, and thus they will be removed last when the virtual space is truncated in the canonical basis. The problem can be solved by using natural spinors which are  optimized\cite{Kumar2017FrozenTheory} for a specific response properties or a different truncation scheme. However, investigating them is outside the scope of the present manuscript. The perturbative correction to the FNS-CCSD and FNS-CCSD(T) energies improves the convergence behaviour of the dipole moment and it converges smoothly with the size of the truncated virtual space in the MP2 corrected FNS-CCSD and FNS-CCSD(T) methods. Neese and co-workers{\cite{Datta2016AnalyticTheory}} have shown that the inclusion of MP2 level correction does not lead to any significant improvement over uncorrected pair natural orbital (PNO) based CCSD dipole moment. One should note that the non-relaxed MP2 density has been used for the perturbation correction in the ref 44. Now, the orbital relaxation plays a major role in determining the accuracy of one-electron properties calculated using MP2 method, unlike in the case of CCSD method where the singles amplitudes already bring a part of the orbital relaxation effect. The perturbational correction in the present finite-field approach corresponds to use of a relaxed MP2 density matrix, which can better reproduce the behaviour of CCSD dipole moment. 
Figure \ref{fig:my_label4} presents the convergence of FNS-CCSD and FNS-CCSD(T) dipole moment with respect to the FNS truncation threshold for HF. It can be seen that our chosen default threshold leads to the convergence of FNS-CCSD(T) results for dipole moment when MP2 correction is considered. The errors in FNS-CCSD dipole moment are below 0.001 Debye with the default FNS truncation threshold.  Table \ref{table:2} presents the dipole moment of hydrogen halides and coinage metal hydrides in FNS-CCSD and FNS-CCSD(T) method with an FNS truncation threshold of $10^{-5}$. The error in the MP2 corrected dipole moment is less than 0.003 Debye in both FNS-CCSD and FNS-CCSD(T) methods with respect to their untruncated canonical analog and is less than the error bar of the CCSD(T) method\cite{helgaker2014molecular}. The dipole moment in FNS-CCSD(T) method show good agreement with experimental results for hydrogen halides, except for HI which shows a deviation of 0.05 Debye. The possible source of the error can be incompleteness of the used basis set,  missing core correlation effect due to the frozen core approximation, missing higher order relativistic and higher order correlation effects. A detailed study on the comparison of four-component relativistic CCSD(T) dipole moment with the available experimental results will be followed up in a separate manuscript.

\subsection{Bond Length and Harmonic Vibrational  Frequency}
\begin{figure}[h!]
    \centering
    \includegraphics[scale=0.5]{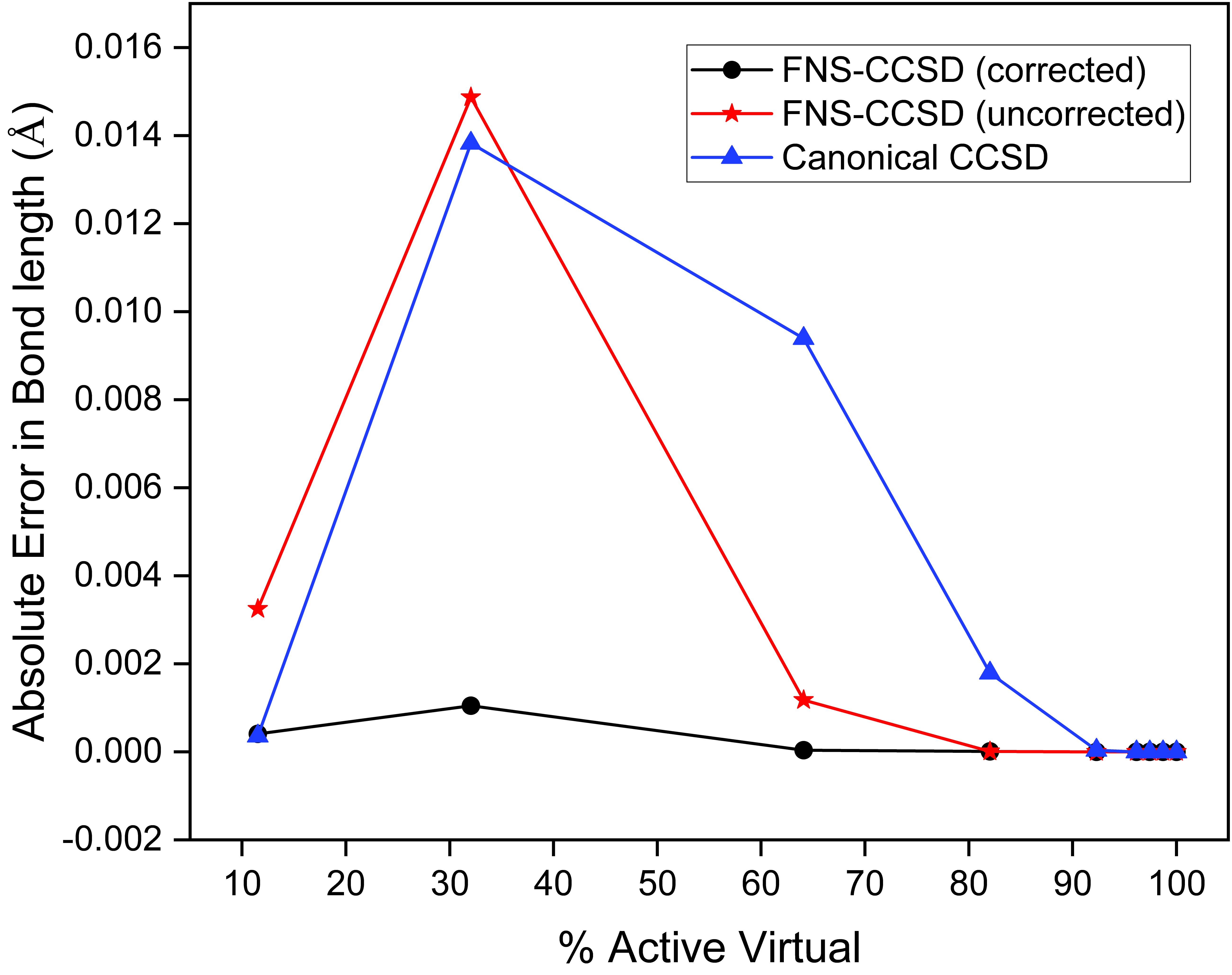}\\
 \caption{ The convergence of the four-component relativistic CCSD bond length with respect to the size of the virtual space in canonical and FNS basis for HF.}
    \label{fig:my_label5}
\end{figure}

\begin{figure}[h!]
    \centering
    \includegraphics[scale=0.5]{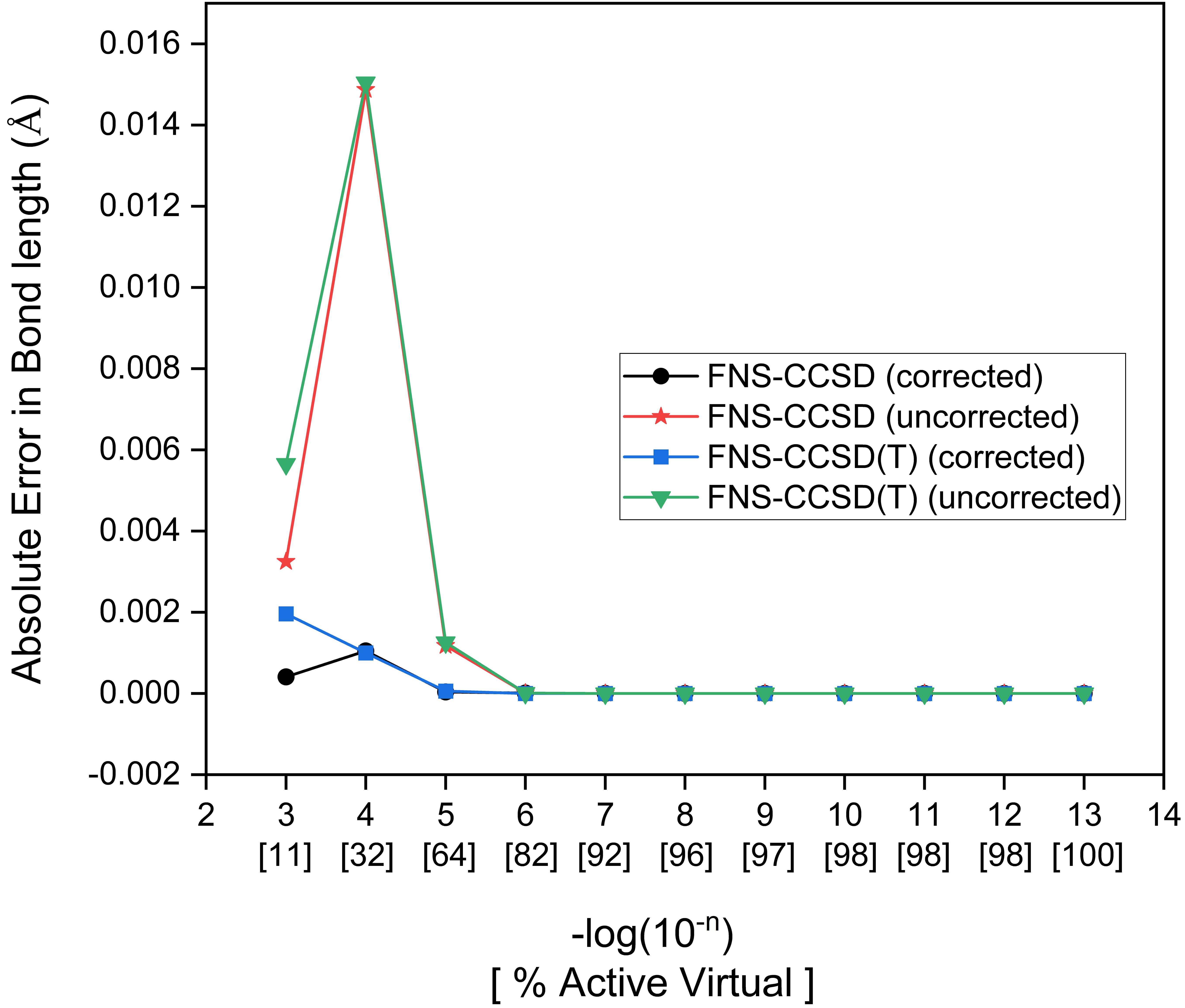}\\
 \caption{ The convergence of FNS-CCSD and FNS-CCSD(T) bond length with respect to the FNS truncation threshold  ($10^{-n}$) for HF. The percentage of the kept virtual space is provided in the brackets.}
    \label{fig:my_label6}
\end{figure}

\begin{figure}[h!]
    \centering
    \includegraphics[scale=0.5]{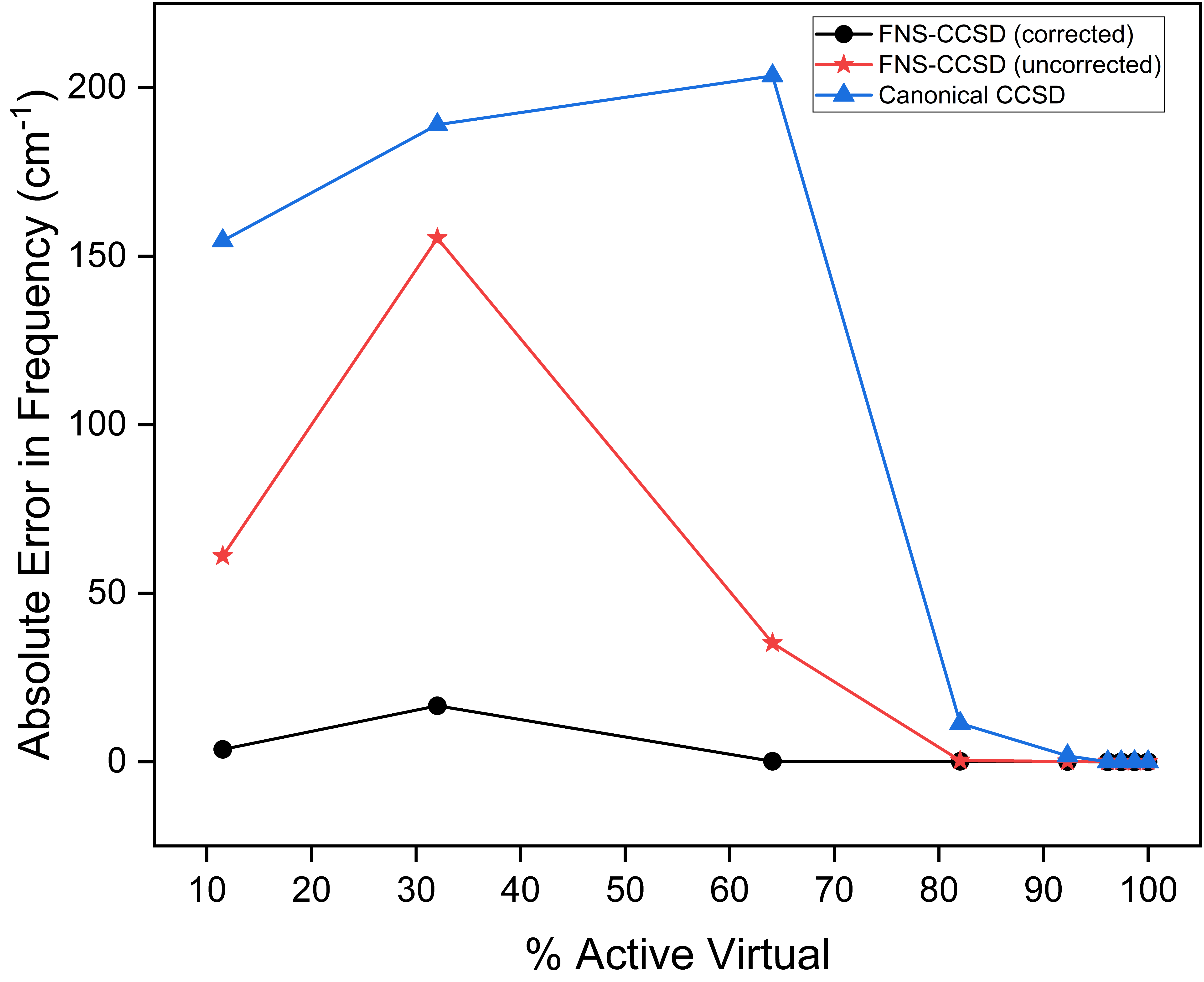}\\
 \caption{  The convergence of the four-component relativistic CCSD harmonic vibrational frequency with respect to the size of the virtual space in canonical and FNS basis for HF.}
    \label{fig:my_label7}
\end{figure}

\begin{figure}[h!]
    \centering
    \includegraphics[scale=0.5]{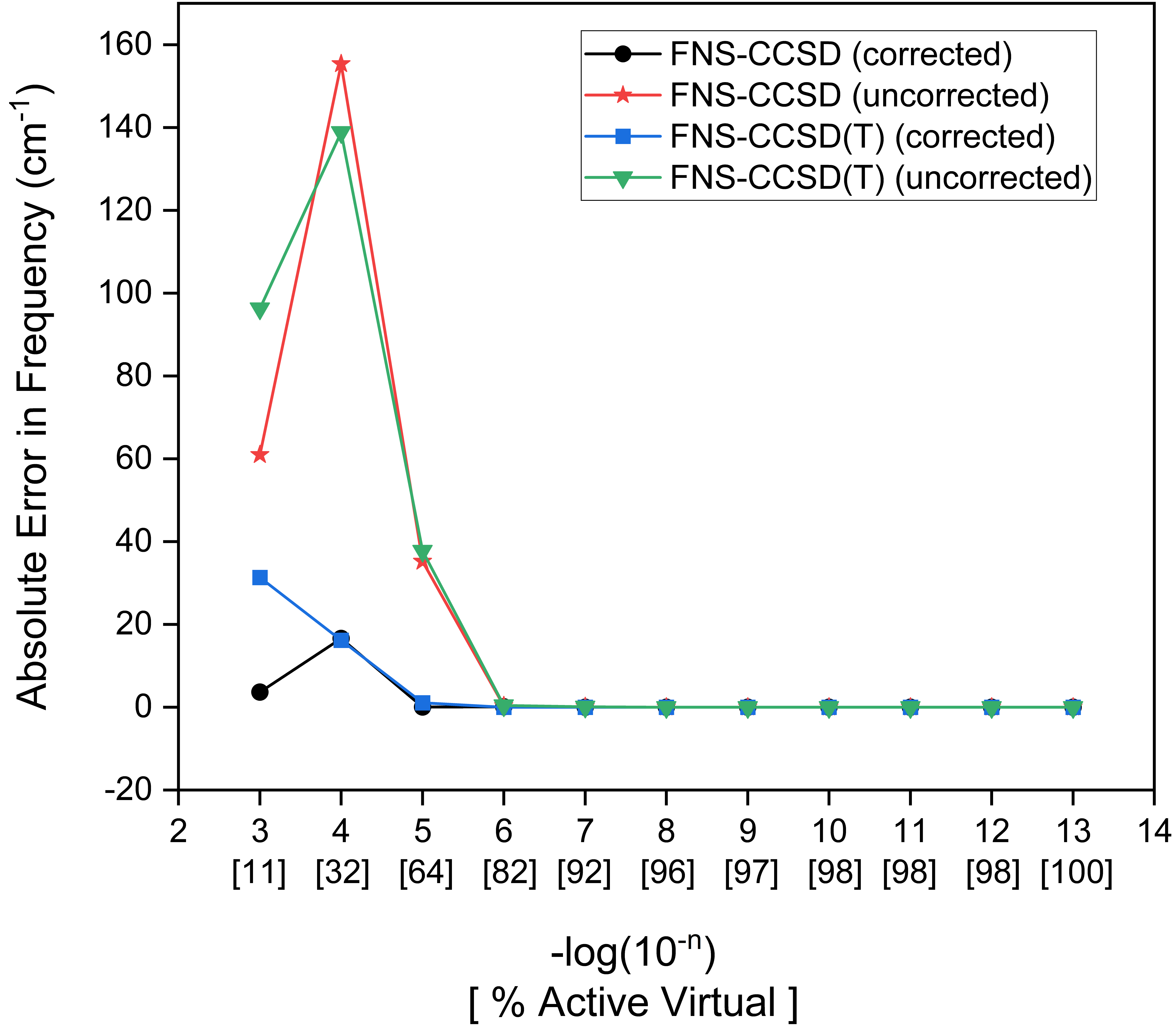}\\
 \caption{ The convergence of FNS-CCSD and FNS-CCSD(T) harmonic vibrational frequency with respect to the FNS truncation threshold ($10^{-n}$) for HF. The percentage of the kept virtual space is provided in the brackets.}
    \label{fig:my_label8}
\end{figure}

We have calculated the bond length and harmonic vibrational frequency in FNS-CCSD and FNS-CCSD(T) methods by numerical differentiation of the total energy using the TWOFIT utility program of DIRAC19\cite{DIRAC21}. A fifth-degree polynomial was used and the CCSD energies were converged up to ${10^{-12}}$. Figure \ref{fig:my_label5} presents the convergence of the bond length of HF in CCSD method with respect to the size of the virtual space in canonical and FNS basis. The convergence of the bond length with respect to the size of the virtual space is quite erratic in both canonical and FNS basis. However, the convergence is more rapid in the FNS basis. The inclusion of MP2 correction smoothens out the convergence behavior to a great extent and the bond length converges with $\sim$ 65\% of the total natural virtual spinor space.  Similar trend is observed for FNS-CCSD(T) method (see Figure S3). Figure \ref{fig:my_label6} shows the convergence of FNS-CCSD and FNS-CCSD(T) bond length with respect to the FNS threshold for HF. The threshold of $10^{-5 }$ is sufficient to get converged results for bond length of HF. Table \ref{table:3} presents the bond length of hydrogen halides and coinage metal hydrides with an FNS threshold of $10^{-5 }$. The maximum error in uncorrected FNS-CCSD(T) method with respect to its canonical analog is less than 0.0041 \AA, which is one order of magnitude less than the error bar of CCSD(T) method\cite{helgaker2014molecular} with respect to experiments. The error further reduces with the MP2 correction. In the present case the FNS-CCSD(T) results are within 1 picometer of experimental values. 
\\ The harmonic vibrational frequencies show trends very similar to the bond length. The convergence of the harmonic vibrational frequencies with the size of the virtual space in CCSD method is erratic in both FNS and canonical basis (See Figure \ref{fig:my_label7}). However, the convergence  is faster in the FNS basis and the use of MP2 correction helps to achieve a smooth convergence. Similar trend is observed in FNS-CCSD(T) method (see Figure S4). From Figure \ref{fig:my_label8} it can be seen that a threshold of $10^{-5 }$ is good enough to get convergence in FNS-CCSD and FNS-CCSD(T) harmonic vibrational frequency results for HF, when MP2 correction is considered. Table \ref{table:4} presents the  harmonic frequency of hydrogen halides and coinage metal hydrides with the FNS threshold of $10^{-5 }$. The harmonic frequency results for all the molecules in FNS-CCSD and FNS-CCSD(T) method are within 10 $cm^{-1}$ of their canonical analog, when MP2 correction is considered. The calculated FNS-CCSD(T) results show excellent agreement with the available experimental results for harmonic vibrational frequencies.

\section{Conclusions}
We present the theory, implementation and the benchmarking of an efficient four-component relativistic coupled cluster method based on frozen natural spinors. The correlation energy converges more rapidly with respect to the size of the virtual space in the FNS basis than that observed in the canonical basis. We have used the criteria of the natural spinor occupation threshold to truncate the virtual space and the convergence of the correlation energy is smooth with respect to the truncation threshold. However, the same trend does not hold for property calculations. The convergence of the error for dipole moment is slower in FNS basis than that observed in the canonical spinor basis. The error in the bond length and harmonic vibrational frequencies converges faster in the FNS basis. However the convergence is not smooth with respect to the truncation parameters. The inclusion of second-order perturbative correction drastically reduces the magnitudes of error in both energy and property calculations. In addition, the perturbative correction makes the convergence of properties smooth with respect to the truncation threshold. A truncation threshold of $10^{-5}$ provides uniform accuracy over molecules even though the size of the virtual space is different from molecule to molecule in FNS basis. This makes the four-component relativistic coupled cluster calculations based on FNS basis black box to use.

The use of frozen natural spinors can drastically reduce the computational cost of four-component relativistic coupled cluster calculations for energy and properties without any significant loss in accuracy. However, the routine use of the FNS based coupled cluster method will require more efficient implementation, elaborate benchmarking, implementation of analytic property calculations, and extension to excited states. Work is in progress towards that direction. 
\section*{Supporting Information}
The Supporting Information is available. The number of the frozen occupied and virtual spinors used for the calculations, additional benchmarking results for FNS-CCSD(T) method, all the Dirac-Fock, canonical CCSD, canonical CCSD(T), FNS-CCSD, FNS-CCSD(T) energies are provided in the supporting information.
\section*{Acknowledgment}
The authors acknowledge the support from the IIT Bombay, IIT Bombay Seed Grant project, DST-SERB, CSIR-India, DST-Inspire Faculty Fellowship for financial support, IIT Bombay super computational facility, and C-DAC Supercomputing resources (PARAM Yuva-II, Param Bramha) for computational time.

\section*{Conflict of interest}
The authors declare no competing financial interest.

\newpage
\bibliography{references,main}

\newpage
\newpage
\newpage
\begin{table}
\centering
\arrayrulecolor{black}
\Huge
\caption{The percentage of the correlation energy recovered in FNS-CCSD and FNS-CCSD(T) method with an FNS truncation threshold of $10^{-5}$.}
\resizebox{17cm}{!}{%
\begin{tabular}{!{\color{black}\vrule}l!{\color{black}\vrule}l!{\color{black}\vrule}l!{\color{black}\vrule}l!{\color{black}\vrule}l!{\color{black}\vrule}l!{\color{black}\vrule}l!{\color{black}\vrule}} 
\hline
\multicolumn{7}{!{\color{black}\vrule}l!{\color{black}\vrule}}{\hspace{30cm}\textbf{ Hydrogen Halides}~}                                                                                                                                                                                                                                                                 \\ 
\hline
\multirow{2}{*}{Molecule~} & \multirow{2}{*}{Total Virtual Spinors~} & \multirow{2}{*}{Active Virtual Spinors~} & \multicolumn{4}{l!{\color{black}\vrule}}{\hspace{12cm}\% Correlation Energy Recovered~}                                                                                                                                                                \\ 
\cline{4-7}
                           &                                         &                                          & FNS-CCSD (corrected)~ & \begin{tabular}[c]{@{}l@{}}FNS-CCSD (uncorrected)~\\~\end{tabular} & \begin{tabular}[c]{@{}l@{}}FNS-CCSD(T) (corrected)~\\~\end{tabular} & \begin{tabular}[c]{@{}l@{}}FNS-CCSD(T) (uncorrected)~\\~\end{tabular}  \\ 
\hline
HF~                        & 156~                                    & 100~                                     & 100.08~               & 98.46~                                                             & 100.04~                                                             & 98.45~                                                                 \\ 
\hline
HCl~                       & 182~                                    & 114~                                     & 99.99~                & 95.21~                                                             & 99.96~                                                              & 95.30~                                                                 \\ 
\hline
HBr~                       & 362~                                    & 178~                                     & 100.14~               & 99.04~                                                             & 100.09~                                                             & 99.02~                                                                 \\ 
\hline
HI~                        & 434~                                    & 174~                                     & 100.13~               & 99.32~                                                             & 100.09~                                                             & 99.30~                                                                 \\ 
\hline
\multicolumn{7}{!{\color{black}\vrule}l!{\color{black}\vrule}}{\hspace{30cm}\textbf{ Coinage Metal Hydrides}~}                                                                                                                                                                                                                                                           \\ 
\hline
CuH~                       & 210~                                    & 182~                                     & 99.99~                & 99.94~                                                             & 99.99~                                                              & 99.93~                                                                 \\ 
\hline
AgH~                       & 224~                                    & 174~                                     & 100.01~               & 99.81~                                                             & 99.97~                                                              & 99.77~                                                                 \\ 
\hline
AuH~                       & 236~                                    & 206~                                     & 100.00~               & 99.96~                                                             & 100.00~                                                             & 99.96~                                                                 \\
\hline
\end{tabular}%
}
\label{table:1}
\arrayrulecolor{black}
\end{table}
\clearpage
\newpage
\newpage
\begin{table}
\centering
\arrayrulecolor{black}
\Huge
\caption{The dipole moment (Debye) in FNS-CCSD and FNS-CCSD(T) method with an FNS truncation threshold of $10^{-5}$}
\resizebox{17cm}{!}{%
\begin{tabular}{!{\color{black}\vrule}l!{\color{black}\vrule}l!{\color{black}\vrule}l!{\color{black}\vrule}l!{\color{black}\vrule}l!{\color{black}\vrule}l!{\color{black}\vrule}l!{\color{black}\vrule}l!{\color{black}\vrule}} 
\hline
\multicolumn{8}{!{\color{black}\vrule}l!{\color{black}\vrule}}{\hspace{30cm}\textbf{Hydrogen Halides}~}                                                                                                                                                                                          \\ 
\hline
\multirow{2}{*}{Molecule~} & \multirow{2}{*}{Canonical CCSD~} & \multicolumn{2}{l!{\color{black}\vrule}}{Error in Dipole Moment~} & \multirow{2}{*}{Canonical CCSD(T)~} & \multicolumn{2}{l!{\color{black}\vrule}}{Error in Dipole Moment~} & \multirow{2}{*}{Experimental Value\cite{nelson1967selected}~}  \\ 
\cline{3-4}\cline{6-7}
                           &                                  & FNS-CCSD (corrected)~ & FNS-CCSD (uncorrected)~                   &                                     & FNS-CCSD(T) (corrected)~ & FNS-CCSD (uncorrected)~                &                                       \\ 
\hline
HF~                        & 1.8055~                          & -0.0010~              & 0.0302~                                   & 1.7878~                             & 0.0001~                  & 0.0314~                                & 1.82~                                 \\ 
\hline
HCl~                       & 1.0826~                          & -0.0012~              & 0.0236~                                   & 1.0702~                             & -0.0006~                 & 0.0242~                                & 1.08~                                 \\ 
\hline
HBr~                       & 0.7937~                          & 0.0020~               & -0.0170~                                  & 0.7805~                             & 0.0010~                  & -0.018~                                & 0.82~                                 \\ 
\hline
HI~                        & 0.4003~                          & -0.0017~              & 0.0~                                      & 0.3889~                             & -0.0016~                 & 0.0001~                                & 0.44~                                 \\ 
\hline
\multicolumn{8}{!{\color{black}\vrule}l!{\color{black}\vrule}}{\hspace{30cm}\textbf{Coinage Metal Hydrides}~}                                                                                                                                                                                    \\ 
\hline
CuH~                       & 2.9295~                          & 0.0014~               & -0.0025~                                  & 2.6964~                             & 0.0028~                  & -0.0011~                               & ~                                     \\ 
\hline
AgH~                       & 3.0469~                          & 0.0022~               & 0.0013~                                   & 2.8209~                             & 0.013~                   & 0.0003~                                & ~                                     \\ 
\hline
AuH~                       & 1.4439~                          & -0.0026~              & 0.0131~                                   & 1.2565~                             & -0.0016~                 & 0.0142~                                & ~                                     \\
\hline
\end{tabular}%
}
\label{table:2}
\arrayrulecolor{black}
\end{table}
\clearpage
\newpage
\newpage
\begin{table}
\centering
\arrayrulecolor{black}
\Huge
\caption{The bond length (\AA) in FNS-CCSD and FNS-CCSD(T) method with an FNS truncation threshold of $10^{-5}$.}
\resizebox{17cm}{!}{%
\begin{tabular}{!{\color{black}\vrule}l!{\color{black}\vrule}l!{\color{black}\vrule}l!{\color{black}\vrule}l!{\color{black}\vrule}l!{\color{black}\vrule}l!{\color{black}\vrule}l!{\color{black}\vrule}l!{\color{black}\vrule}} 
\hline
\multicolumn{8}{!{\color{black}\vrule}l!{\color{black}\vrule}}{\hspace{30cm}\textbf{Hydrogen Halides}~}                                                                                                                                                                                      \\ 
\hline
\multirow{2}{*}{Molecule~} & \multirow{2}{*}{Canonical CCSD~} & \multicolumn{2}{l!{\color{black}\vrule}}{Error in Bond Length~} & \multirow{2}{*}{Canonical CCSD(T)~} & \multicolumn{2}{l!{\color{black}\vrule}}{Error in Bond Length~} & \multirow{2}{*}{Experimental Value\cite{huber2013molecular}~}  \\ 
\cline{3-4}\cline{6-7}
                           &                                  & FNS-CCSD (corrected)~ & FNS-CCSD (uncorrected)~                 &                                     & FNS-CCSD(T) (corrected)~ & FNS-CCSD(T) (uncorrected)~           &                                       \\ 
\hline
HF~                        & 0.9183 ~                         & ~0.0 ~               & ~-0.0011 ~                              & ~0.9213~                            & ~0.0 ~                  & ~-0.0012 ~                           & ~0.9168~                              \\ 
\hline
HCl~                       & 1.2753 ~                         & 0.0~                 & -0.0022~                                & 1.2777~                             & 0.0~                    & -0.0022~                             & 1.2745~                               \\ 
\hline
HBr~                       & 1.4089~                          & 0.0005~               & -0.0038~                                & 1.4117~                             & 0.0003~                  & -0.0040~                             & 1.4144~                               \\ 
\hline
HI~                        & 1.6027~                          & 0.0004~               & -0.0004~                                & 1.6059~                             & 0.0004~                  & -0.0005~                             & 1.6091~                               \\ 
\hline
\multicolumn{8}{!{\color{black}\vrule}l!{\color{black}\vrule}}{\hspace{30cm}\textbf{Coinage Metal Hydrides}~}                                                                                                                                                                                \\ 
\hline
CuH~                       & 1.4689~                          & -0.0009~              & 0.0005~                                 & 1.4581~                             & -0.0011~                 & 0.0007~                              & 1.4626~                               \\ 
\hline
AgH~                       & 1.6218~                          & 0.0~                 & -0.0017~                                & 1.6130~                             & -0.0004~                 & -0.0021~                             & 1.618~                                \\ 
\hline
AuH~                       & 1.5188~                          & 0.0~                  & -0.0001~                                & 1.5170~                             & 0.0~                    & -0.0001~                             & 1.5238~                               \\
\hline
\end{tabular}%
}
\label{table:3}
\arrayrulecolor{black}
\end{table}
\clearpage
\newpage

\newpage
\begin{table}
\centering
\arrayrulecolor{black}
\Huge
\caption{The harmonic vibrational frequency ($cm^{-1}$) in FNS-CCSD and FNS-CCSD(T) method with an FNS truncation threshold of $10^{-5}$.}
\resizebox{17cm}{!}{%
\begin{tabular}{!{\color{black}\vrule}l!{\color{black}\vrule}l!{\color{black}\vrule}l!{\color{black}\vrule}l!{\color{black}\vrule}l!{\color{black}\vrule}l!{\color{black}\vrule}l!{\color{black}\vrule}l!{\color{black}\vrule}} 
\hline
\multicolumn{8}{!{\color{black}\vrule}l!{\color{black}\vrule}}{\hspace{30cm}\textbf{Hydrogen Halides}~}                                                                                                                                                                                  \\ 
\hline
\multirow{2}{*}{Molecule~} & \multirow{2}{*}{Canonical CCSD~} & \multicolumn{2}{l!{\color{black}\vrule}}{Error in Frequency~} & \multirow{2}{*}{Canonical CCSD(T)~} & \multicolumn{2}{l!{\color{black}\vrule}}{Error in Frequency~} & \multirow{2}{*}{Experimental Value~\cite{huber2013molecular}}  \\ 
\cline{3-4}\cline{6-7}
                           &                                  & FNS-CCSD (corrected)~ & FNS-CCSD (uncorrected)~               &                                     & FNS-CCSD(T) (corrected)~ & FNS-CCSD(T) (uncorrected)~         &                                       \\ 
\hline
HF~                        & 4083.08~                         & -0.08~                & 35.22~                                & 4036.44~                            & 1.06~                    & 37.72~                             & 4138.32~                              \\ 
\hline
HCl~                       & 3015.13~                         & -0.03~                & 9.48~                                 & 2990.66~                            & 0.24~                    & 10.44~                             & 2990.94~                              \\ 
\hline
HBr~                       & 2673.1~                          & -0.63~                & 9.14~                                 & 2647.42~                            & -0.03~                   & 7.99~                              & 2648.97~                              \\ 
\hline
HI~                        & 2333.7~                          & -9.09~                & 17.98~                                & 2306.75~                            & -7.59~                   & 19.93~                             & 2309.01~                              \\ 
\hline
\multicolumn{8}{!{\color{black}\vrule}l!{\color{black}\vrule}}{\hspace{30cm}\textbf{Coinage Metal Hydrides}~}                                                                                                                                                                            \\ 
\hline
CuH~                       & 1898.89~                         & -8.65~                & 27.31~                                & 1939.15~                            & -2.64~                   & 26.56~                             & 1941.26~                              \\ 
\hline
AgH~                       & 1746.12~                         & 2.46~                 & 0.88~                                 & 1762.98~                            & 3.45~                    & 4.79~                              & 1759.9~                               \\ 
\hline
AuH~                       & 2306.83~                         & 6.19~                 & -7.8~                                 & 2317.95~                            & 5.41~                    & -8.3~                              & 2305.01~                              \\
\hline
\end{tabular}%
}
\label{table:4}
\arrayrulecolor{black}
\end{table}

\end{document}